\documentstyle[12pt,aaspp4]{article}

\newcommand{\sso}{$^{-1}$}
\begin{document}
\title{The Mass Inflow Rate in the Barred Galaxy NGC 1530}
\author{Michael W. Regan\altaffilmark{1},\\ 
Stuart N. Vogel\altaffilmark{2},\\
\& \\
Peter J. Teuben\altaffilmark{3}}
\affil{Department of Astronomy,\\ University of Maryland,\\ 
College Park, MD 20742}
\altaffiltext{1}{Email-mregan@astro.umd.edu}
\altaffiltext{2}{Email-vogel@astro.umd.edu}
\altaffiltext{3}{Email-teuben@astro.umd.edu}
\begin{abstract}
Mass inflow in barred galaxies has been invoked to account for
a wide variety of phenomena, but until now direct evidence
for inflow has been lacking.
We present Fabry-Perot H$\alpha$ observations of
the barred spiral galaxy NGC 1530
from which we determine velocities of the ionized gas for
the entire region swept by the bar.
We compare the velocity field to models of gas flow
in barred spirals and show that it is well reproduced by
ideal gas hydrodynamic models.
Inspection of the models and observations reveals that gas
entering the bar dust lanes streams directly down the dust lanes toward the
2 kpc radius nuclear ring.
The models predict that
approximately 20\% of the gas flowing down the dust lane enters the nuclear ring;
the remaining gas sprays around the ring to the other bar dust lane.
The fraction of the gas entering the ring is relatively insensitive to the shape or
size of the bar.
Our observations of the velocity field and dust optical depth 
yield a mass inflow rate into the nuclear ring
of 1 M$_{\sun}$ yr\sso.

\end{abstract}

\section{Introduction}
Mass inflow in galactic bars
is postulated to fuel central starbursts
(\markcite {HS94}{Heller \& Shlosman 1994}) and active galactic nuclei
(\markcite {SFB89}{Shlosman, Frank, \& Begelman 1989}).
Statistical evidence links starbursts with barred galaxies
(\markcite {H96}{Ho 1996}), suggesting that mass inflow may be driven
by bars.
In addition to producing starbursts, such inflow is also proposed to influence
chemical evolution
 (\markcite{RB93}{Roy \& Belley 1993};
\markcite {FBK94}{Friedli, Benz, \& Kennicutt 1994};
\markcite{MR94}{Martin \& Roy 1994}), 
to produce the central concentrations of molecular gas observed in barred
galaxies 
(\markcite {K92}{Kenney et al 1992}), to create bulges in late-type spirals
(\markcite {N96}{Norman et al 1996}), and ultimately to destroy the
bar (\markcite {FB93}{Friedli \& Benz 1993};
\markcite {N96} {Norman et al 1996}).
Although most models of gas flow in barred spirals
predict mass inflow and observations exist of gas with
inward radial motions (\markcite {BSK}{Benedict, Smith, \& Kenney 1996};
\markcite {Q95} {Quillen et al 1995}), observational
evidence of net mass inflow is lacking.

By studying the gas distribution and
kinematics in barred spirals it should be possible to estimate
the mass inflow rate into the nuclear region.
The inflow could be measured if density-weighted galactic
radial velocities could be found at all azimuth angles,
 assuming steady-state flow (Athanassoula 1992; hereafter A92).
However, only the line-of-sight velocity can be measured,
precluding a model-independent confirmation of mass inflow.
On the other hand, if the gas flow could be correctly modeled, the mass inflow
could be determined by fitting a model to the data.

Although many properties of the ISM
needed to model gas flow 
in a galaxy are known,
current numerical models do not incorporate all 
that is known due to a lack of computing power and understanding
of how these properties change in various environments.
Since no single model can model all the relevant physical
processes and size scales,
a wide range of numerical representations for the gas have been used,
including grid-based ideal-gas hydrodynamics 
(\markcite {A92}{A92};
\markcite {PST95}{Piner, Stone \& Teuben 1995; hereafter PST95}), 
smooth-particle-hydrodynamics (SPH)
(\markcite {WH92}{Wada \& Habe 1995};
\markcite{FB93}{Friedli \& Benz 1993}),
and massless sticky particles (clouds)
(\markcite{CG85}{Combes \& Gerin 1985}; 
\markcite{B94}{Byrd et al 1994}).
Even the physical mechanism for the loss of angular momentum by the gas
is controversial.
The gravitational torque exerted by the stellar bar on the gas in the 
offset dust lanes has
been postulated as the cause of the loss of angular momentum 
by the gas
(\markcite {van81}{van Albada \& Roberts 1981}; 
\markcite{CG85}{Combes \& Gerin 1985}; 
\markcite{C88}{Combes 1988};
\markcite{SN93}{Shlosman \& Noguchi 1993};
\markcite{Q95}{Quillen et al 1995};
\markcite{C96}{Combes 1996}).
Others propose that
gas is driven inward when it loses angular momentum 
from hydrodynamic torques when it is
shocked in the dust lanes along the leading 
edge of the bar 
(\markcite{A92}{A92}; 
\markcite {PST95}{PST95}).
A mass inflow rate was calculated for NGC 7479 based on the offset
between the major axis of the stellar potential and the molecular gas
assuming that the resulting gravitational torque on the gas drives
the gas inward
(\markcite {Q95}{Quillen et al 1995}).
A detailed comparison of the various sources of torque will
be presented in Regan (1997). 
Even though
the different models predict the 
same basic morphology for the dust and gas in
barred spirals, they differ in the 
predicted mass inflow rates.
Significantly, they also make different predictions for the kinematics
of the gas. 
Thus, it should be possible to determine the model that best simulates the
true ISM with
detailed kinematic observations.

Our inspection of the gas hydrodynamic models 
(\markcite{A92}{A92};
\markcite{PST95}{PST95})
shows that all gas that encounters the bar dust lanes flows down
the dust lane to the nuclear region.
Not all of the gas entering the nuclear region remains there since
some of the gas sprays back into the bar region 
at the contact point of the dust lane and the nuclear ring 
(\markcite{B91}{Binney et al 1991}).
If the models are correct, direct measurement of the inflow
rate is straightforward: simply measure the mass and velocity of gas in 
the dust lane and correct for the amount of gas that flows back into
the bar region.
The first step is to determine whether the hydrodynamic models provide a good
description of the gas flow, and to do this we compare our complete
velocity field with predictions of the hydrodynamic models.
Although we cannot compare the velocity fields of other models 
because
only the hydrodynamic models provide detailed and 
complete velocity fields, we
expect that other models that have short mean-free-paths for the gas
encountering the dust lane will also fit the data and predict that
most gas encountering the dust lane will flow directly to the
nuclear ring.

This paper is the third in a series that looks at the barred spiral galaxy
NGC 1530.
In our first paper (\markcite {RVT95}{Regan, Vogel, \& Teuben 1995}; hereafter Paper I) we 
discussed the molecular gas and dust morphology of NGC 1530.
In our second paper (\markcite {RTVV96}{Regan et al 1996}; hereafter Paper II)
we discussed the morphology of the stars, atomic gas, and ionized gas and
determined the rotation curve based on HI and H$\alpha$ observations.
In this paper we will discuss the kinematics of the ionized gas
and show that the hydrodynamic models
are a good fit to the observations.
We also compare the kinematics of the ionized gas in the
dust lanes to Berkeley-Illinois-Maryland-Association millimeter array
(BIMA) observations of the kinematics of the molecular gas.
From the observations we will then make an estimate of the mass inflow rate
into the nuclear ring.

\section{Fabry-Perot Observations of H$\alpha$ Emission}

\begin{figure}[htbp!]
\plotfiddle {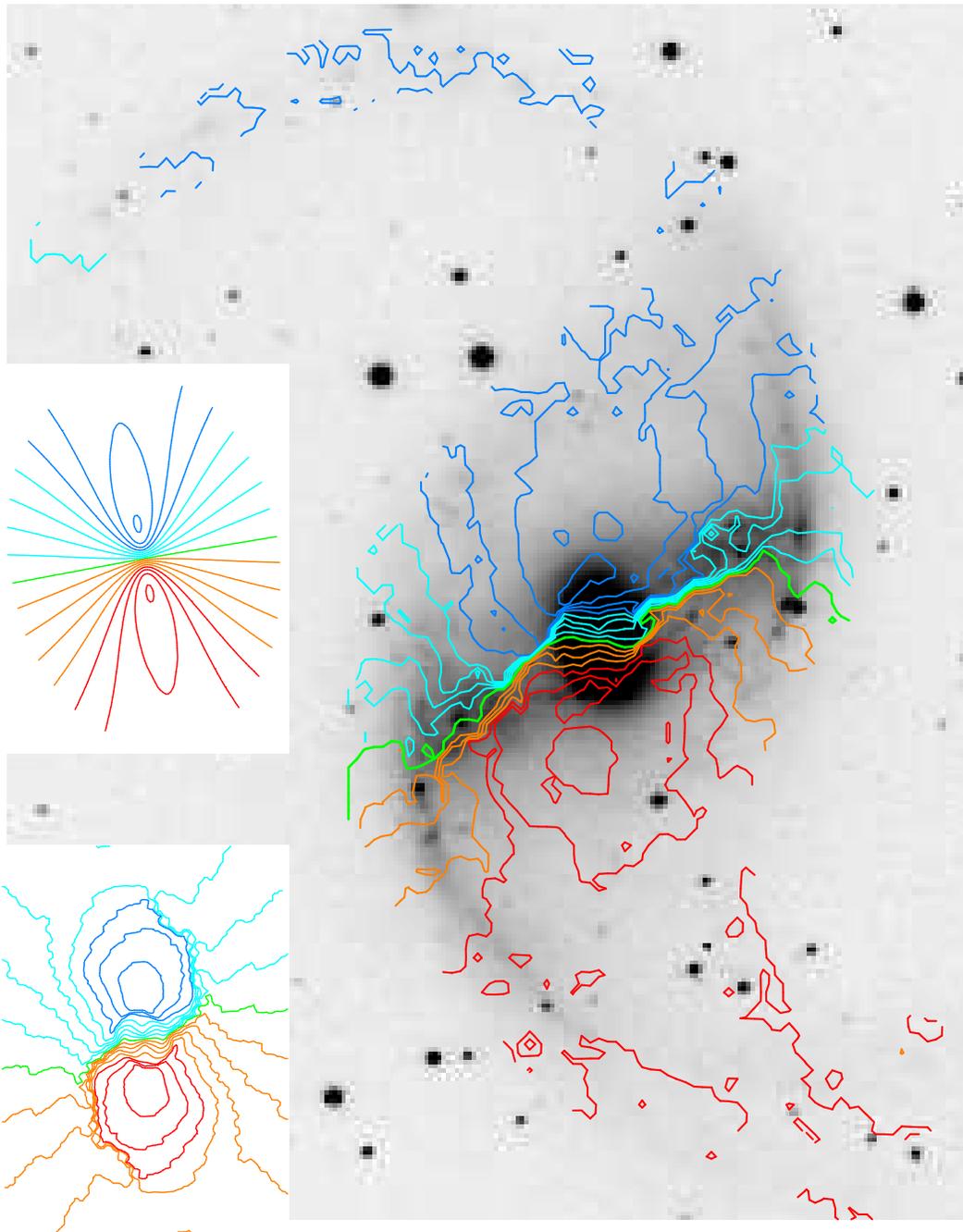} {6.4in} {0.} {72} {72} {-200} {-45}
\caption[H$\alpha$ velocity field of NGC 1530 compared to barred and 
unbarred model galaxies]{ The observed H$\alpha$ velocity field
of NGC 1530 compared to model velocity fields.
The observed isovelocity
contours are overlaid on an I-band image.
The contours range from 2250 to 2650 km s\sso\ at 25 km s\sso\ intervals
with the lower velocities being in the northern half of the galaxy.
(Lower left insert) A model velocity field for a barred galaxy
from PST95 projected to the same
orientation on the sky as NGC 1530.
Note the good agreement between the model and actual isovelocity contours.
(Middle left insert)
Velocity field for a model with the same radial mass distribution but with
no bar. }
\label{modelnvel}
\end{figure}

We observed H$\alpha$ emission from 
NGC 1530 on the 30 September and 1 October 1994 
using the
Maryland-Caltech Imaging Fabry-Perot Interferometer on the 1.5m at Palomar
Observatory.
The observations have an angular resolution that ranged 
from 4\arcsec\ where the signal is strong to
15\arcsec\ in fainter regions. The
velocity resolution is 25 km s\sso; 
profiles can be centroided to a 1 $\sigma$ accuracy
of 2 km s\sso\ in strong emission regions and
10 km s\sso\ in fainter regions.
Our 3 $\sigma$
sensitivity limit is 5 x $10^{-18}$erg cm$^{-2}$ s$^{-1}$ arcsec$^{-2}$
(emission measure 2.5 cm$^{-6}$ pc for T$_e$=10$^4$K).
The details of the data reduction process were described in Paper II.
The resulting
velocity field obtained using a moment technique is shown in
Figure \ref{modelnvel}.
The velocity field was determined by using a variable resolution method
that results in higher resolution where there is more signal.
We detect emission over most of the region inside a radius of 1\farcm5 
(15 kpc);
this includes the entire region swept by the bar.

\section{Results}

\subsection{Comparison with Models}

In Figure \ref{modelnvel} we compare the observed NGC 1530 velocity field
with a model hydrodynamic velocity field (Model 4) from PST95;
other recent grid-based models of gas flow in barred spirals 
(e.g. \markcite {A92}{A92})
give similar predictions for the gas streamlines for the region swept by
the bar dust lane.
Also shown is the velocity field corresponding to a model galaxy
with the same radial mass distribution but without a bar.
The bar model has an axial ratio of 2.5, a bar quadrupole moment
of 4.5 x 10$^{10}$ M\sun\ kpc$^2$, corotation at 1.2 times the
bar radius, and is a 
reasonable match to NGC 1530.
Comparison of the barred and unbarred models
shows that the effects of the bar on the isovelocity contours include:
1) significant compression of the contours coinciding with the bar dust lane
(a jump of typically 200 km s\sso\ in the line of sight) ending at
the inner Lindblad resonance (ILR) ring,
2) rotation of the kinematic major axis by $\sim$20\arcdeg\ clockwise,
3) an apparent boost in the rotation speed along an axis perpendicular
to the bar,
4) twisting of the contours in the bulge region becoming 
uniformly spaced and parallel
as the ILR radius is approached, and 
5) a shock along the gaseous spiral arms.
All of these features are seen in the observed velocity field of NGC 1530.
Two features in the model not apparent in the
observations, the pinching of the isovelocity contours near the nucleus and
the decline in the rotational velocities along the major axis, are probably
due to differences between the mass distribution of the model and the 
actual mass distribution in NGC 1530.
Another difference is that the shock in the spiral arms is not
as strong in the observations as it is in the model,
perhaps due to the absence of a spiral potential in the model.
In general, there is excellent agreement considering that 
the PST95 model was not generated for NGC 1530 and there was no 
tuning of the model.
The only parameters adjusted to fit the NGC 1530 velocity field were
the inclination, position angle of the bar,
position angle of the galaxy, and size scale.
Presumably even better agreement could be obtained by adjusting the 
mass model, bar shape, and bar strength to match 
the observations.

It is hard for the particle-based models (SPH and sticky particles) to
create a full 2D velocity field because in regions of low 
density there are very few particles and thus the velocity field is
poorly sampled.
Combes and Gerin (\markcite{cg85}{1985}) did publish an isovelocity 
diagram for the central region of their sticky-particle model galaxy.
The isovelocity contours of NGC 1530 do not 
support
the model of Combes and Gerin (\markcite{cg85}{1985}) since the
model does not show the twist in the contours at the 
ILR nor the compression
of contours along the leading edge of the bar.
In addition, the dust lane morphology in their model does not 
match the dust lane morphology of NGC 1530 (\markcite{RVT95}{Paper I})
since the model dust lanes continue down the leading edge of
the bar past the nucleus while the dust lanes in NGC 1530 end at the
nuclear ring.
One SPH study that does give a velocity field for the bar region of a weak
bar is
Wada \& Habe (\markcite{wh95}{1995}).
The dust lane morphology of their models and their velocity fields
are very similar in the dust lanes although they do not form
as strong of a nuclear ring as the PST95 models.
Thus, we conclude that 
both grid-based and SPH ideal gas models are better matches to the data
than the published cloud-based sticky particle models.

It is possible that particle-based models with other parameters might
fit our observations.  
The narrow width of the dust lane and the
associated large velocity gradient can be used to constrain the particle
mean-free-path.  
Clearly, the mean free path of the gas responsible for
the dust lane extinction must be shorter than the width of the dust
lane.  
The gas traced by H$\alpha$ must have a similar short mean free
path, since we observe large velocity gradients associated with the dust
lane.  
If clouds punched through the dust lanes, downstream from the
dust lanes clouds would be expected to have a range of velocities,
because the clouds would not adjust to the post-shock velocities as fast
as the diffuse gas. 
Since the observed line widths remain relatively narrow ($\sim$50 km s\sso)
downstream of the dust lanes, this argues that if clouds punch through
the dust lanes, they cannot contribute significantly to H$\alpha$
emission.  
We conclude that a model which reproduces the dust lanes and
H$\alpha$ velocity field must have a cloud mean-free-path less than 300
pc (2\arcsec). 

In summary, the grid-based ideal gas models match the 
kinematic observations better than any particle-based models with published
kinematics.
However, the best comparison would be to use the same mass model for the
two methods.

\subsection{Determination of the Mass Inflow Rate}

Examination of the gas flow 
in the rotating reference frame of the hydrodynamic models 
(\markcite {A92}{A92}; \markcite {PST95}{PST95})
reveals that 
all the gas that enters the nuclear ring arrives via the dust lanes\footnote
{Note that Figure 2b is incorrect in PST95. 
The correct figure
shows gas streamlines similar to \markcite {A92}{(A92)}.}.
The general flow of gas in the bar region is that
gas encounters the dust lanes in a shock, and is redirected down
the dust lane toward the nuclear ring.
At the contact point of the
dust lane and the nuclear ring there is a region of
divergence, sometimes referred to as the spray shock
(\markcite{B91}{Binney et al 1991});
here some of the gas enters the nuclear ring, and some of it sprays around
to encounter the other dust lane.
This flow pattern is distinctly different from the more common view that clouds
cross the dust lanes, slowly spiraling into the nuclear region (\markcite {C96}{Combes 1996}).

The mass flux along the dust lane is an upper limit to the mass
inflow rate, and that flux can be calculated as follows. 
The mass flux, $\dot{M}$, at any distance, $d$, along the dust lane
can be expressed as
\begin{equation}
\dot{M}(d) = \sigma(d)~ W~ V_{dl}(d)
\end{equation}
where $\sigma$ is the  gas mass surface density, $W$ is the width
of the dust lane, and $V_{dl}$ is the velocity of the gas along the dust lane.
We determine $\sigma$ and $W$ from BVRIJHK observations (Paper I).
At a resolution of 4\arcsec\, the range in values for $\sigma$ is 
$15 < \sigma < 40$ M$\sun$ pc$^{-2}$, while
$ W = 0.9$ kpc.

The dust lane velocity, $V_{dl}$, is determined from the
Fabry-Perot observations
because $V_{dl}$ is parallel to the dust lane.  
Several steps are needed
to obtain $V_{dl}$ from the observed velocity of the gas, $V_{obs}$, 
at some position $d$ in the dust lane with
a projected galactocentric 
radius of $R^\prime$ at a projected angle of $\theta^\prime$ from
the minor axis of the galaxy.
The unprojected radius, $R$, can be expressed as:
$
R(d)=R^\prime(d)(\cos^2\theta(d) +  \sin^2\theta(d)  \cos^{-2}i)^{1/2},
$
where $\theta = \tan^{-1}(\tan\theta^\prime\cos i)$ is the true
angle relative to the minor axis, and
$i=45\arcdeg$ is the inclination of NGC 1530.

To obtain $V_{dl}(d)$,  $V_{obs}(d)$
must be corrected for the various projections, 
all of which are known, and also the projected
pattern speed, $V_{pat}(d)= \omega_p R(d) \sin\theta(d) \sin i$, where
$\omega_{p}$ is the angular pattern speed.
The angular pattern speed was determined in Paper II to be
20 km s\sso\ kpc\sso.  
Therefore, 
$V_{dl}(d)=(V_{obs}(d)-V_{sys}-V_{pat}(d))(\cos\gamma\cos\theta(d) 
+\sin\gamma\sin\theta(d))^{-1}\sin i^{-1},
$
where $\gamma=45$\arcdeg\ is the angle the dust lane makes with the minor axis
 and
$V_{sys}$ is the systemic velocity.
As predicted, V$_{dl}$ is observed to be directed toward the nuclear ring in
both dust lanes.  We find that $ 85 < | V_{dl} | < 200$ km s\sso\ and is
generally increasing in magnitude towards the nuclear ring.

\begin{figure}[ht!]
\plotfiddle {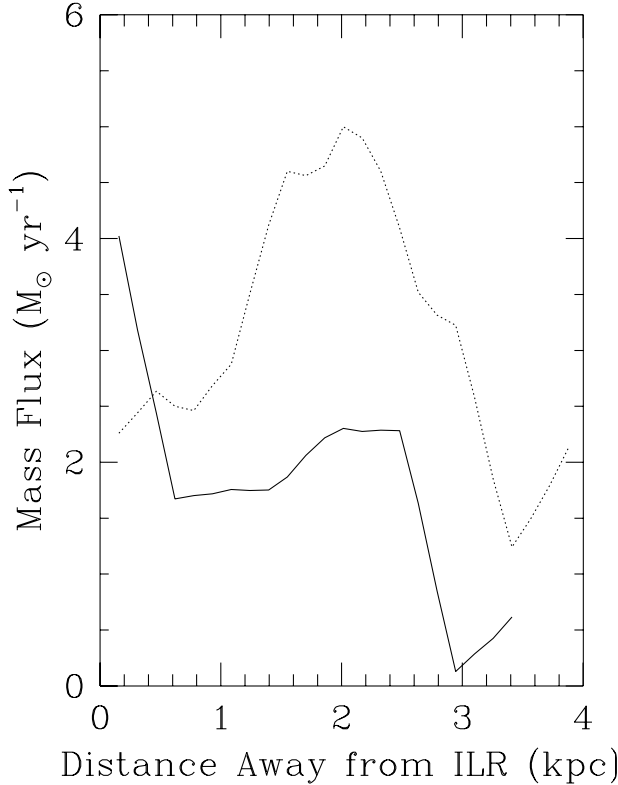} {4.0in} {0.} {40} {40} {-150} {0}
\caption[Mass flux in the dust lanes of NGC 1530]
{Mass flux in the dust lanes versus distance from the nuclear ring.
The western dust lane data are plotted as a solid line and 
the eastern dust lane data as a dotted line. 
There is a clear trend in the western dust lane
consistent with the mass flux in the dust lane increasing toward
the nuclear ring. The trend is not seen in the eastern 
dust lane.}
\label{dustlength}
\end{figure}

\begin{figure}[ht!]
\plotfiddle {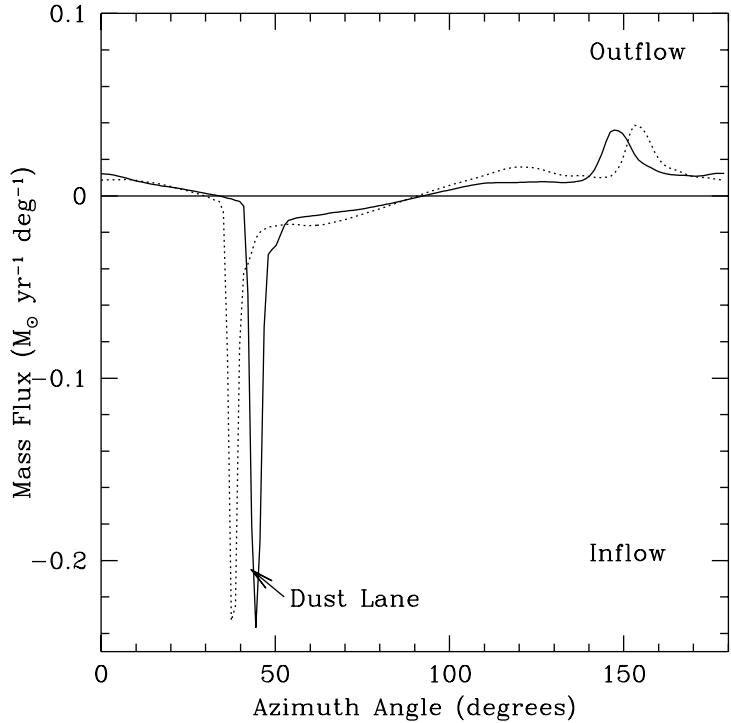} {3.5in} {0.} {50} {50} {-150} {-80}
\caption[Mass inflow as a function of angle]
{Mass inflow as a function of angle over 1/2 of the bar
derived from hydrodynamic models. The two lines are at different radii. The
solid line is at about 1.2 times the radius of the ILR ring and the dashed
line is at 1.3 times the radius of the ILR ring. Note that the primary
dust lanes along the leading side of the bar
dominate the mass inflow but that there is significant outflow in a broad
dust lane on the trailing side of the bar.
Also, note that the net inflow rate in this model is 
not the same as in our observations.}
\label{inflow}
\end{figure}

Using equation (1) and the observed values of $\sigma$, $W$, and $V_{dl}$, 
we
calculated $\dot{M}$(d) for each dust lane (Figure \ref{dustlength}).
The
combined mass inflow in the two dust lanes is
6 $\pm$3 M$_{\sun}$ yr\sso\ which is an upper limit to the overall mass inflow
rate.
To show that the dust lane mass flux overestimates the overall mass
flux we calculated the mass flux
as a function of angle using a snapshot of the time dependent models
of PST95
(Figure \ref{inflow}).
The dominance of the dust lane is clearly revealed but the gas flowing
out at other locations also affects the net mass flux.
By running a grid of models of varying bar parameters 
(axial ratio, bar quadrupole moment, central mass concentration) we
find that the mass flux in the dust lane ranges from three to seven times
the overall mass flux into the ILR ring.
Using this range and the observed dust lane mass flux
yields a net flux
into the ILR ring of 1$^{+2}_{-0.5}$ M$_{\sun}$ yr\sso. 
For a uniform injection rate of gas into the barred region,
$\dot{M}$ is expected to increase toward the ILR ring 
(i.e. with decreasing $d$) since
mass is added to the dust lane all along its length and remains in the dust
lane until it reaches the ILR ring.  
Variations in the injection rate could explain the absence of a clear
monotonic increase in $\dot{M}$ toward the ILR.  However, there are significant
uncertainties in the measured quantities which make conclusions about
variations premature.  
Our resolution of 4\arcsec\ is larger than the width of the dust lanes,
which leads to uncertainties in $V_{dl}$ since it
changes by 40-50 km s\sso\ from pixel to pixel across
the dust lane.
Also, H$\alpha$ is not a linear tracer and V$_{obs}$ can be
biased by asymmetric ionizing flux.
In addition, the dust mass determination has uncertainties
caused by star formation and uncertainties in the dust-free colors.
All of these uncertainties can be greatly reduced with higher resolution
observations of the velocities and extinction.

A key assumption in our derivation of $\dot{M}$ 
is that velocities of the ionized
gas accurately measure the velocities of the gas whose mass is determined
from extinction observations in the dust lane.  
This assumption can
be tested in the dust lanes close to the ILR ring, where CO emission is
detected in the BIMA maps of CO 1$-$0 emission.  
We find that here the
H$\alpha$ and CO velocities agree to within 15$-$30 km s\sso.  
This difference
is small compared to $V_{dl}$ = 100 km s\sso, 
which implies that H$\alpha$ velocities are a
reasonably reliable tracer of the velocities of the
high column density gas.
The observed velocities are also consistent with observations of inflowing CO
velocities at the terminus of the dust lanes of NGC 4314 ($\sim$80 km\sso)
(\markcite {BSK96}{Benedict et al 1996}). 

As we have shown, in published hydrodynamic models all the gas in bar
dust lanes flows directly along the dust lane toward the ILR ring.  The situation
in other models is less clear.  Models with gas mean-free-paths that are long
compared to the width of the dust lane obviously will give a different
answer; however, these models do not account for the observed velocity field
or narrow dust lanes and are inappropriate for understanding gas flow in the
dust lanes.  It is likely that particle models with short
mean-free-paths could also be consistent with the observations and would also
predict gas flow along the dust lanes to the ILR ring (J.M. Stone, personal
communication).

\section{Giant Molecular Clouds in the Bar Region}

The excellent agreement between the observations and the ideal gas models
implies that the gas traced by the H$\alpha$ kinematics is relatively
diffuse and not concentrated in giant molecular clouds (GMCs).
This is consistent with the small amount of CO detected along the
bar (Paper I; Downes et al 1996).
Locally, the majority of the mass in the ISM is contained within 
GMCs (e.g. \markcite {B78}{Blitz 1978}).
It may be that processes unique to a strongly barred galaxy either quickly
destroy GMCs or inhibit their formation within the radius swept by the bar.
The unique gas flow in the barred region suggests a mechanism for either
process.  
The gas flow exhibits large
divergence in the streamlines prior to the dust lanes, which may
tear apart GMCs (thought not to be strongly gravitationally
bound) or prevent formation of GMCs.  
Another possibility is that GMCs 
will have their lower column density regions stripped each time they pass
through the dust lanes.
If some GMCs survive and make their way into the nuclear ring
by another path, our estimate of the mass inflow rate would 
be a lower limit.

\section{Evolution of Bars}

The observed mass inflow rate in NGC 1530 could have implications for
the long term evolution of the bar.
A problem in both analytical and n-body simulations of the long term
stability of bars in the presence of a halo is that the bar slows
down in several rotation times 
due to angular momentum transfer from the bar to the 
halo (\markcite {W85}{Weinberg 1985}).
Only if angular momentum is added to the bar can it remain
rotating with corotation near the bar end, as it is in NGC 1530 (Paper II).
Since the angular momentum lost by the gas as it moves
from the bar end to the nuclear ring is gained by the stars in the
bar, we can estimate the torque provided to the bar.
Using our inflow rate of 1 M$_{\sun}$ yr\sso, a bar radius of 10 kpc,
a nuclear ring radius of 2 kpc, and a rotational speed of 220 km s\sso, we
derive a torque of 2$\times$10$^3$ M$_{\sun}$ km s\sso\ kpc yr\sso.
This torque adds enough angular momentum to
double the angular momentum of
the bar in approximately 5 Gyr or approximately one fifth
the rate at which the bar loses angular momentum to the halo.
Therefore, this inflow rate does not provide enough angular momentum to
the bar to offset that lost to the halo.

Using the dust extinction data, the mass in the dust lanes is 
$1.8\times10^8$ M$_{\sun}$.
In the hydrodynamic models approximately one-third of the gas in the region
swept by the dust lanes is in the dust lanes.  Since gas flows
into the ILR ring at 1 M$_{\sun}$ yr\sso, this implies that, 
absent replenishment
from outside, the bar region will deplete its gas in  5 $\times 10^8$ yr.  

Using the same method employed to estimate the mass of the dust lanes, the
ILR ring has a mass of 2.3 $\times 10^8$ M$_{\sun}$.  
The present accretion rate of 1
M$_{\sun}$ yr\sso\ is approximately equal to the star formation rate
in the central region of the galaxy (Paper II).

\section{Conclusions}

We have shown that the velocity field of NGC 1530 
obtained from H$\alpha$ Fabry-Perot observations
and the gas morphology inferred from CO and dust extinction data are 
in excellent
agreement with the predictions of ideal gas models.
In addition, the observed narrow H$\alpha$ line widths downstream of the shock,
the increasing mass in the dust lane as it nears the nuclear ring,
and the narrow 
width of the dust lanes all show that if there is a significant
component of mass in GMCs they must have short mean-free-paths.
The gas streamlines are strongly affected by the hydrodynamic forces
leading to a radically different view of gas flow than models that
ignore hydrodynamic forces.
Examination of the hydrodynamic models reveals
that gas that enters the
dust lane streams toward the nuclear ring directly along the dust lane.
Not all of the gas in the dust lane enters the nuclear ring since 
the models reveal that some of
it renters the bar region at the contact point of the dust lane and the
nuclear ring.
Using observations of the extinction in the dust lane to determine
the mass, Fabry-Perot observations to obtain the
velocity, and the average ratio of dust lane mass flux to overall mass
flux we have derived
a mass inflow rate into the nuclear ring of NGC 1530 of 
1$^{+2}_{-0.5}$ M$_{\sun}$ yr\sso.

\acknowledgements
We would like to thank Jeff Kenney, 
Jim Stone, Alice Quillen, and David Spergel
 for helpful discussions and comments.
This work was supported in part by NSF grant 
AST 9314847.

\clearpage

\end{document}